\documentclass[conference]{IEEEtran}
\IEEEoverridecommandlockouts

\usepackage{amsmath,amssymb,amsfonts}
\usepackage{graphicx}
\usepackage{textcomp}
\usepackage{xcolor}
\usepackage{cleveref}
\usepackage{cite}

\def\BibTeX{{\rm B\kern-.05em{\sc i\kern-.025em b}\kern-.08em
	T\kern-.1667em\lower.7ex\hbox{E}\kern-.125emX}}
	
\usepackage{tikz}

\newcommand\copyrighttext{%
	\footnotesize \textcopyright 2020 IEEE. Personal use of this material is permitted.
	Permission from IEEE must be obtained for all other uses, in any current or future
	media, including reprinting/republishing this material for advertising or promotional
	purposes, creating new collective works, for resale or redistribution to servers or
	lists, or reuse of any copyrighted component of this work in other works.
}
\newcommand\copyrightnotice{%
	\begin{tikzpicture}[remember picture,overlay]
	\node[anchor=south,yshift=10pt] at (current page.south) {\fbox{\parbox{\dimexpr\textwidth-\fboxsep-\fboxrule\relax}{\copyrighttext}}};
	\end{tikzpicture}%
}

\begin{document}

\title{Adaptive Transmission Parameters Selection Algorithm for URLLC Traffic in Uplink  
	\thanks{The research has been carried out at IITP RAS and supported by the grant No 18-37-20077 mol-a-ved of the Russian Foundation for Basic Research.}
}

\author{
	\IEEEauthorblockN{
		Aleksei Shahsin\IEEEauthorrefmark{1}\IEEEauthorrefmark{2}, Andrey Belogaev\IEEEauthorrefmark{1}, Artem Krasilov\IEEEauthorrefmark{1}\IEEEauthorrefmark{2} and Evgeny Khorov\IEEEauthorrefmark{1}\IEEEauthorrefmark{2}
	}

	\IEEEauthorblockN{\IEEEauthorrefmark{1}Institute for Information Transmission Problems, Russian Academy of Sciences, Moscow, Russia}
	\IEEEauthorblockN{\IEEEauthorrefmark{2}Moscow Instutute of Physics and Technology, Moscow, Russia}
	\IEEEauthorblockN{e-mail: \{shashin, belogaev, krasilov, khorov\}@wireless.iitp.ru}
}

\maketitle

\copyrightnotice

\begin{abstract}
	Ultra-Reliable Low-Latency Communications (URLLC) is a novel feature of 5G cellular systems. To satisfy strict URLLC requirements for uplink data transmission, the specifications of 5G systems introduce the grant-free channel access method. According to this method, a User Equipment (UE) performs packet transmission without requesting  channel resources from a base station (gNB). With the grant-free channel access, the gNB configures the uplink transmission parameters in a long-term time scale. Since the channel quality can significantly change in time and frequency domains, the gNB should select robust transmission parameters to satisfy the URLLC requirements. Many existing studies consider fixed robust uplink transmission parameter selection that allows satisfying the requirements even for UEs with poor channel conditions.  However, the more robust transmission parameters are selected, the lower is the network capacity. In this paper, we propose an adaptive algorithm that selects the transmission parameters depending on the channel quality based on the signal-to-noise ratio statistics analysis at the gNB. Simulation results obtained with NS-3 show that the algorithm allows meeting the URLLC latency and reliability requirements while reducing the channel resource consumption more than twice in comparison with the fixed transmission parameters selection.
\end{abstract}

\begin{IEEEkeywords}
	5G, URLLC, Grant-free, Uplink, MCS, K-repetition, parameters selection
\end{IEEEkeywords}

\section{Introduction}
Ultra-Reliable Low-Latency Communications (URLLC) is a novel traffic type that will be supported by the next-generation cellular networks (5G) in addition to enhanced Mobile Broadband (eMBB) and massive Machine-Type Communications (mMTC)~\cite{3gpp.38.913}. Many applications, such as autonomous vehicles interaction or telesurgery, generate uplink URLLC-traffic~\cite{urllc_scenarios}. In particular, each autonomous vehicle transmits information collected from its sensors, e.g., information about its position, speed, acceleration, or obstacles detected on the road. Similarly, during the telesurgery, a surgeon transmits commands to remote robotic manipulators that react to these commands with corresponding actions. To ensure the efficient work of such applications, they impose very strict latency (several milliseconds) and reliability (higher than $99.999\%$) requirements~\cite{vision2015framework}. It should be noted that problem of quality of service provisioning for these applications is not only considered for cellular networks, but also attracts much attention from the wired and Wi-Fi communities~\cite{be_rta}. In this paper, we focus on solution for 5G cellular networks.

The 3rd Generation Partnership Project (3GPP) specifications, e.g.,~\cite{3gpp.38.214}, for 5G networks describes two methods for an uplink channel access: grant-based and grant-free (configured grant). According to the grand-based method, to access an uplink channel, a user equipment (UE) transmits a scheduling request and waits for a scheduling grant from a base station (called gNB). When the grant is received, the UE starts transmitting data. As a result, the grant-based channel access delays the actual transmission, which makes this method inapplicable to most of URLLC use cases. Hence, the grant-free channel access is usually considered for URLLC data transmission.

With the grant-free channel access, a gNB selects the channel resources and the transmission parameters, e.g., Modulation and Coding Scheme (MCS), for each UE in a long-term time scale. To improve reliability, UEs can perform multiple transmission attempts. Before the transmission attempt, a UE can either wait for the feedback from a gNB related to the previous attempts or perform the transmission without waiting for feedback. In this paper, we consider the scheme without feedback called K-repetition, which implies that a UE makes $K$ transmission attempts for each data packet using the parameters configured by the gNB. Since the K-repetition does not require receiving feedback from the gNB after each transmission attempt, it allows significantly reducing the data transmission latency. To improve the reliability, the gNB uses Hybrid Automatic Repeat reQuest (HARQ) scheme, e.g., Chase Combining (CC)~\cite{chase1985code}.
Many works compare K-repetition with other schemes that use the feedback from a gNB using analytical models~\cite{reliability-analys-gf, analyzing-gf-access} and simulations~\cite{ul-gf-random-access, system-level-analysis}. Their results show that K-repetition is more effective in terms of transmission latency and reliability except for the cases of high load with massive overlapping of transmissions, i.e., when transmissions corresponding to different UEs with high probability use the same channel resources.

Since the channel quality can significantly change in time and frequency domains, and the gNB selects the transmission parameters for relatively long periods of time, a UE should use robust transmission parameters to satisfy strict URLLC quality of service requirements. In this case, the minimization of channel resource consumption is challenging because the more robust parameters are used, the more channels resources are consumed, which leads to lower network capacity, i.e., the number of data flows that can be served simultaneously.
Studies~\cite{reliability-analys-gf, analyzing-gf-access, ul-gf-random-access, system-level-analysis} consider usage of fixed MCS and the number of transmission attempts, so they do not consider the dynamic parameter selection problem.
The papers~\cite{on-capacity,contention-based} studies the selection of $K$ to maximize the number of users in the network. They consider that in the case when the same resources are assigned to multiple UEs, simultaneous transmissions of different UEs may lead to unsuccessful transmissions. However, in both works, the probability of unsuccessful transmission is estimated without taking into account that the probability of unsuccessful transmission depends on the MCS. 

The channel resource allocation problem for K-repetition is considered in~\cite{system-level-k-rep}. The authors propose to divide available channel resources into multiple groups and randomly select a group for each transmission attempt. They propose to select MCS corresponding to the number of these groups (higher number of groups corresponds to higher code rates). This study shows that the K-repetitions scheme can provide the required packet loss probability with the considered in the paper transmission parameters values. However, the authors do not provide any adaptive method for selecting these parameters, e.g., depending on the channel quality.
The paper~\cite{resource_allocation} provides the algorithm for selecting $K$ and the channel resources for transmission. However, this work considers fixed probabilities of successful decoding without taking into account that they depend on the used MCS.
The paper~\cite{adaptive-repetition-control} provides the number of transmission attempts (i.e., $K$) selection algorithm based on an estimation of fading correlation function. The authors suggest using two attempts in the case of low correlation and four in the case of high correlation. Numerical results show that this method can increase reliability and resource efficiency in a multi-user scenario compared to using a fixed number of attempts. However, this work does not consider the problem of selecting MCS and proposes to use a fixed robust MCS.

The provided above literature analysis show that the existing papers do not provide any adaptive method for selecting MCS depending on channel conditions. In this paper, we propose the adaptive transmission parameters (i.e., MCS and number of transmission attempts) selection algorithm based on estimating of the packet loss ratio for each parameter configuration using the Signal-to-Noise Ratio (SNR) statistics at the gNB. 

The rest of the paper is organized as follows. In Section~\ref{sec:alorithm}, we introduce the proposed algorithm. In Section~\ref{sec:perfomance}, we evaluate its preference  with the NS-3 simulator. Finally, Section~\ref{sec:conclusion} concludes the paper.

\section{Proposed algorithm}
\label{sec:alorithm}
In this paper, we propose the transmission parameters selection algorithm that allows a gNB to select MCS and the number of transmission attempts for a UE transmitting URLLC packets in  uplink. 

Let us consider a UE served by a gNB. The gNB configures the parameters for the UE, i.e., the resources that are available for its uplink transmissions and the transmission parameters, i.e., the MCS and the number of transmission attempts $K$. To change the parameters configuration at the UE, the gNB transmits to the UE the configuration messages that include the new MCS and the number of transmission attempts. In its turn, the UE updates the configuration upon the reception of the new parameters. The time-frequency resources assigned to the UE are divided into multiple groups called Resource Block Groups (RBGs). When the UE transmits a packet, it selects an appropriate number of RBGs corresponding to the selected MCS and the packet size. To avoid correlated errors in the consecutive RBGs, we assume that the UE selects RBGs uniformly spaced in the frequency domain. These RBGs are selected independently for each attempt.
The time slots for repeated attempts are also selected uniformly spaced within the packet delay budget to avoid correlated errors in time. 

Initially, after the connection establishment, the gNB configures the most robust uplink transmission parameters (MCS 0 and $K_{max}$, where $K_{max}$ is the maximum number of transmission attempts) because there is no actual SNR statistics for this UE. Based on Sounding Reference Signals (SRS) periodically transmitted by the UE, the gNB estimates the average SNR for each RBG. Then, for each combination of the transmission parameters MCS and $K$ ($K = 1, \text{\dots}, K_{max}$), the gNB estimates the BLock Error Rate (BLER), i.e., the error probability of a single transmission attempt, as follows:
\begin{itemize}
	\item The gNB calculates the number of RBGs $M_\text{MCS}$ that is required for transmission using the considered MCS. Here we assume that the gNB has information about the packet size (e.g, this information can be provided via cross-layer interaction between the gNB and the URLLC application~\cite{xstream}). Then, the gNB randomly selects $M_\text{MCS}$ RBGs that are uniformly spaced in the available bandwidth.
	\item We assume that the total power remains the same regardless of the number of used RBGs since the UE uses the whole power to transmit data. Hence, the gNB recalculates SNR according to
	\vspace{-0.1cm}
	\[ \text{SNR}_\text{MCS}^{(i)} = \frac{M}{M_\text{MCS}} \cdot \text{SNR}_\text{measured}^{(i)}, \vspace{-0.2cm}\]
	where $\text{SNR}_\text{measured}^{(i)}$ is a measured SNR in RBG $i$, $M$ is a total number of RBGs in which the UE transmits SRS.
	\item The gNB uses the Exponential Effective Signal-to-noise ratio Mapping (EESM)~\cite{eesm} error model for BLER estimation.
\end{itemize} 

Let us describe the third step in more detail. According to the EESM model, the gNB maps vector of SNRs for the selected RBGs to a single effective SNR as follows: 
\[ \text{SNR}_\text{eff} = -\beta \ln \left(\frac{1}{|v|}\sum_{n \in v} \exp (-\frac{\text{SNR}_n}{\beta}) \right), \]
where SNR$_n$ is the SNR value in the $n$-th RBG, $v$ is the set of
allocated RBGs, and $\beta$ is a scaling parameter. We use $\beta$ values obtained in~\cite{eesm}. 

We assume that the gNB uses CC to decode several HARQ transmissions. The effective SNR after $q$ transmission attempts is calculated as follows:
\[ \text{SNR}_\text{eff}^{q} = -\beta \ln \left(\frac{1}{|\omega_q|}\sum_{m = 1}^{|\omega_q|} \exp (-\frac{1}{\beta} \sum_{j=1}^{q}\text{SNR}_{m, j}) \right), \]
where $\omega_j = \{RBG^j_1, ..., RBG^j_{|\omega_j|}\}$ is the set of used RBGs for $j$-th transmission attempt (note, that $|\omega_1| = ... = |\omega_q|$), SNR$_{m, j}$ is the SNR experienced in $RBG^j_m$.

\begin{figure}[!t]
	\includegraphics[width=0.5\textwidth]{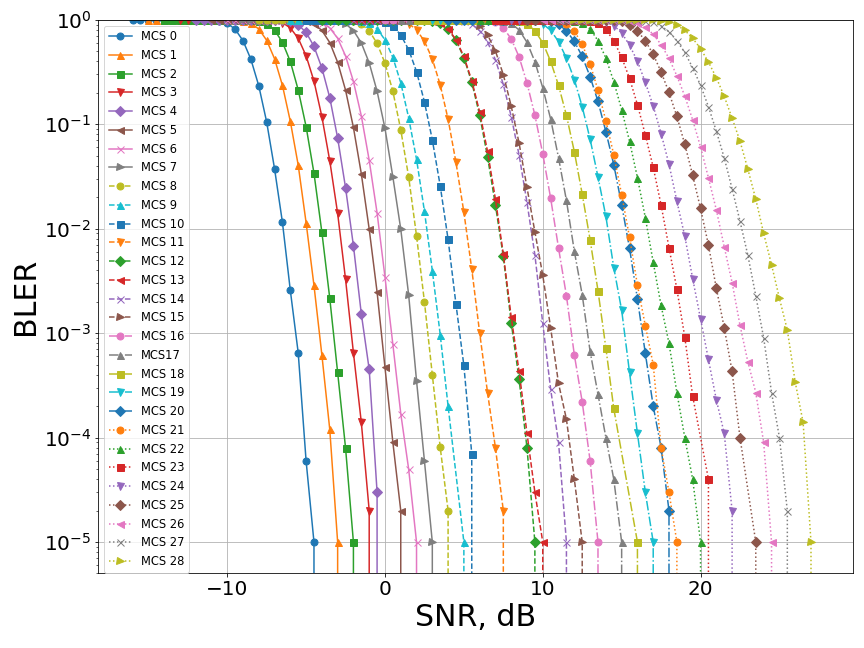}
	\caption{SNR-BLER curves for 25 iterations of a decoder.}
	\label{fig:sinr-bler}
	\vspace{-0.3cm}
\end{figure}

The obtained effective SNR value is mapped to the BLER value using the SNR-BLER curves for the corresponding MCS. In this study, we use SNR-BLER curves (Fig.~\ref{fig:sinr-bler}) obtained for 25 iterations of a decoder for packet size 32 byte and physical layer parameters described in Section~\ref{sec:scenario}. 

The packet loss ratio after all $K$ transmission attempts equals $\text{PLR}_{\text{MCS}, K} = \text{BLER}_1 \cdot \text{\dots} \cdot \text{BLER}_K$, where BLER$_i$ is the obtained BLER for the $i$-th transmission attempt.

As a result, after each SRS reception, the gNB obtains PLR estimations for all the configurations \{MCS, $K$\}. For each configuration, the PLR estimation is averaged with an exponentially weighted moving average as follows:
\[  \widehat{\text{PLR}}_{\text{MCS},K} (t) = \frac{1}{W} \cdot \text{PLR}_{\text{MCS}, K} (t) + (1 - \frac{1}{W}) \cdot \widehat{\text{PLR}}_{\text{MCS}, K} (t-1), \]
where $W$ is the window size, $\widehat{\text{PLR}}_{\text{MCS},K}(t)$ is the averaged PLR estimation after the $t$-th SRS reception.

\begin{figure}[!t]
	\includegraphics[width=0.5\textwidth]{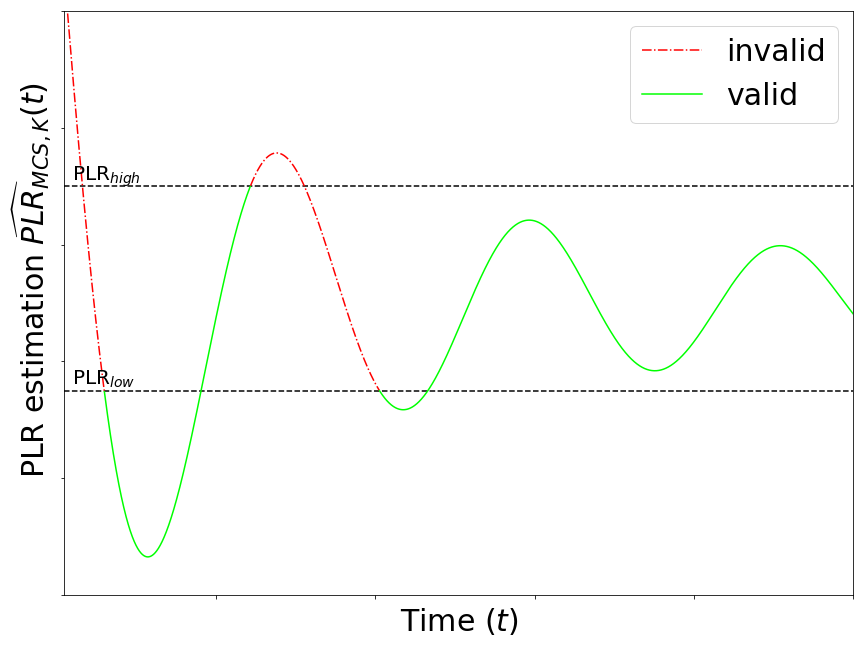}
	\caption{Valid configurations marking.}
	\label{fig:plr-thresholds}
	\vspace{-0.3cm}
\end{figure}

To avoid frequent reconfiguration, i.e., transmission parameters changes at the UE, the gNB uses two thresholds, PLR$_\text{low}$ and PLR$_\text{high}$, where PLR$_\text{low}$ $<$ PLR$_\text{high}$. Specifically, the gNB selects the transmission parameters as follows:
\begin{enumerate}
	\item marks all configurations \{MCS, $K$\} as valid, if $\widehat{\text{PLR}}_{\text{MCS},K}(t)  < \text{PLR}_\text{low}$, and as invalid, if $\widehat{\text{PLR}}_{\text{MCS},K}(t)  > \text{PLR}_\text{high}$ (see Fig.~\ref{fig:plr-thresholds});
	\item calculates the channel resource consumption $M_\text{MCS} \cdot K$ for each valid configuration and selects the one that provides the minimum channel resource consumption.
\end{enumerate}

If the selected configuration differs from the one used by the UE, the gNB sends the new configuration to the UE. 

\section{Performance evaluation}
\label{sec:perfomance}

\subsection{Scenario}
\label{sec:scenario}
We evaluate the performance of the proposed algorithm with the NS-3 simulator~\cite{ns-3}. We consider a single gNB and a single UE that generates Constant Bit Rate (CBR) traffic in the uplink. Specifically, $32$ bytes packets are transmitted with periodicity 10 ms. Each packet should be delivered within the $1$ ms time interval with a probability higher than $99.999\%$. SRSs are transmitted every $5$ ms. 

Let us describe physical layer parameters. Following~\cite{urllc1}, we consider mini-slots (subslots) that consist of two Orthogonal Frequency-Division Multiplexing (OFDM) symbols corresponding to control and data channels. The duration of the OFDM symbol equals $35.7~\mu s$ that corresponds to the $30$ kHz interval between subcarriers. So, there are 14 slots within the packet delay budget.
The bandwidth equals $100$ MHz and consists of $16$ RBGs~\cite{3gpp.38.214}. The UE transmission power equals $23$ dBm, and it is equally distributed between the selected RBGs. We summarize all simulation parameters in Table~\ref{tab:paameters}.


\begin{table}
	\caption{Simulation parameters}
	\label{tab:paameters}
	\begin{tabular}{p{0.223\textwidth} p{0.223\textwidth}}
		\hline
		\textbf{Parameter} & \textbf{Value} \\
		\hline
		Bandwidth & 100 MHz, 16 RBGs \\
		Slot length & $71.4~\mu s$ \\
		Packet size & 32 bytes \\
		Packet period & 10 ms \\
		SRS period & 5 ms \\
		UE power & 23 dBm \\
		UE height & 1.5 m \\
		gNB height & 30 m \\
		Propagation model & Okumura-Hata model~\cite{Okumura} \\
		Fading model & Extended Pedestrian A (EPA)~\cite{fading_models} \\
		Simulation time & 100000 s \\
		$K_{max}$ & 4 \\
		\hline
	\end{tabular}
	\vspace{-0.5cm}
\end{table}

\subsection{Analysis of the results}
\label{sec:results}
As a characteristic of the average channel quality, which decreases with the distance $d$ between the UE and the gNB, we use the value $\text{SNR}_\text{wb}$ called the wideband SNR. This value is calculated as follows: $\text{SNR}_\text{wb} = (P_{TX} \: PL(d)) / (Noise \: BW)$,
where $P_{TX}$ is a  UE transmission power, $BW$ is the available bandwidth, $PL(d)$ is the pathloss, $Noise$ is a noise power spectral density at the gNB. To model the channel quality changes in time and frequency domains, we use the Extended Pedestrian A (EPA) fading model~\cite{fading_models}.

\begin{figure}[t!]
	\includegraphics[width=0.5\textwidth]{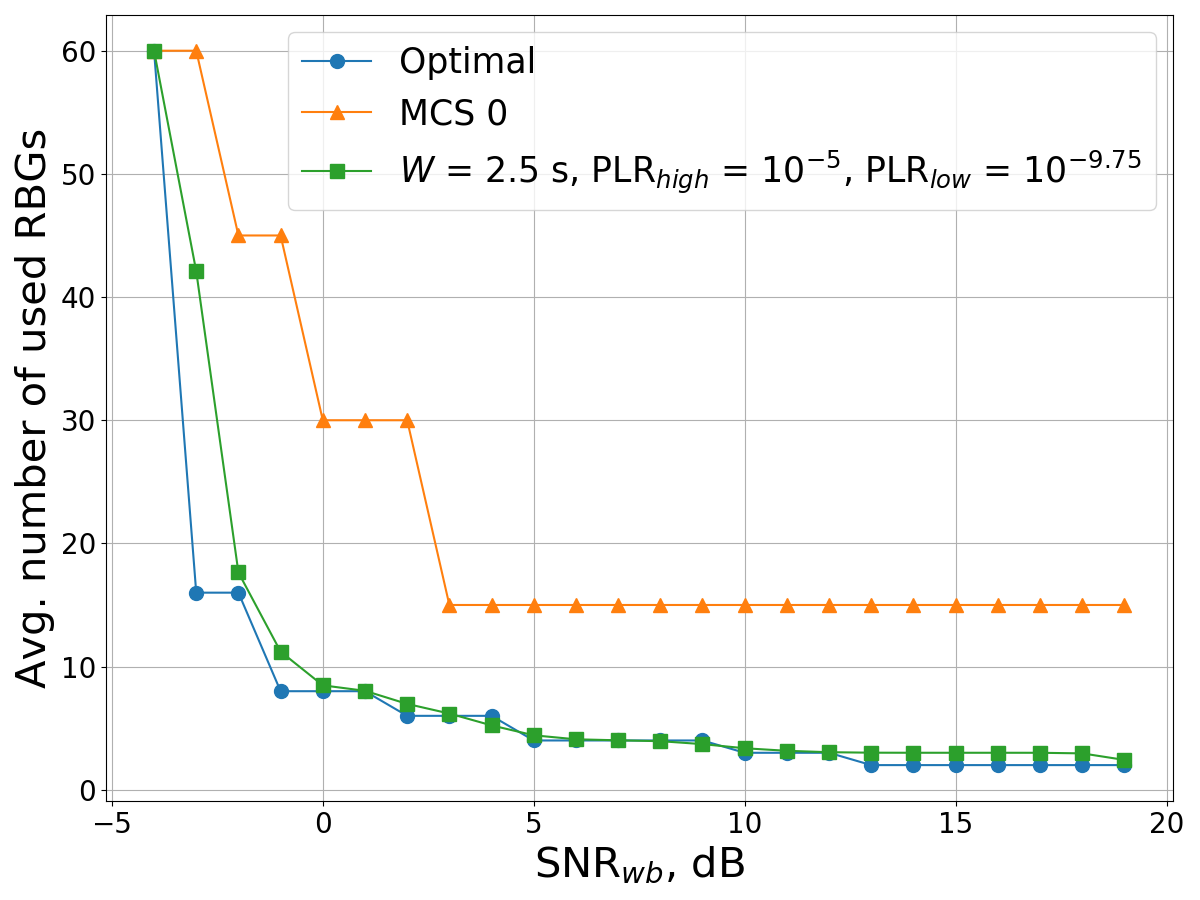}
	\caption{Comparison of fixed and adaptive MCS selection.}
	\label{fig:opt-msc0}
	\vspace{-0.5cm}
\end{figure}

First, let us estimate the reduction of resource consumption that can be achieved with the adaptive transmission parameters selection. For each $\text{SNR}_\text{wb}$ value and each parameter configuration \{MCS, $K$\}, we carry out the full experiment and consider the PLR and the average number of used RBGs obtained in each experiment. Then, for each $\text{SNR}_\text{wb}$, we select the optimal configuration that allows satisfying the URLLC requirements while consuming the minimal number of RBGs. We compare the resource consumption provided by the proposed adaptive algorithm with the resource consumption provided by the optimal configuration and by the fixed selection of the most robust MCS, i.e., MCS 0 (see Fig.~\ref{fig:opt-msc0}). The number of transmission attempts $K \leq K_{max}$ for MCS 0 equals the minimal number that allows satisfying the URLLC requirements. We can see that the proposed algorithm provides more than three times a reduction for the resource consumption in comparison with the MCS 0 selection. Moreover, the results for the proposed algorithm are close to optimal for $\text{SNR}_{\text{wb}} \geq -1$ dB. Since even with the most robust configuration the URLLC requirements cannot be satisfied at $\text{SNR}_{\text{wb}} < -4$ dB, we assume in the further experiments that the UE is located uniformly in the circle with center at gNB and radius corresponding to $\text{SNR}_{\text{wb}} = -4$ dB. 

\begin{figure}[t!]
	\includegraphics[width=0.5\textwidth]{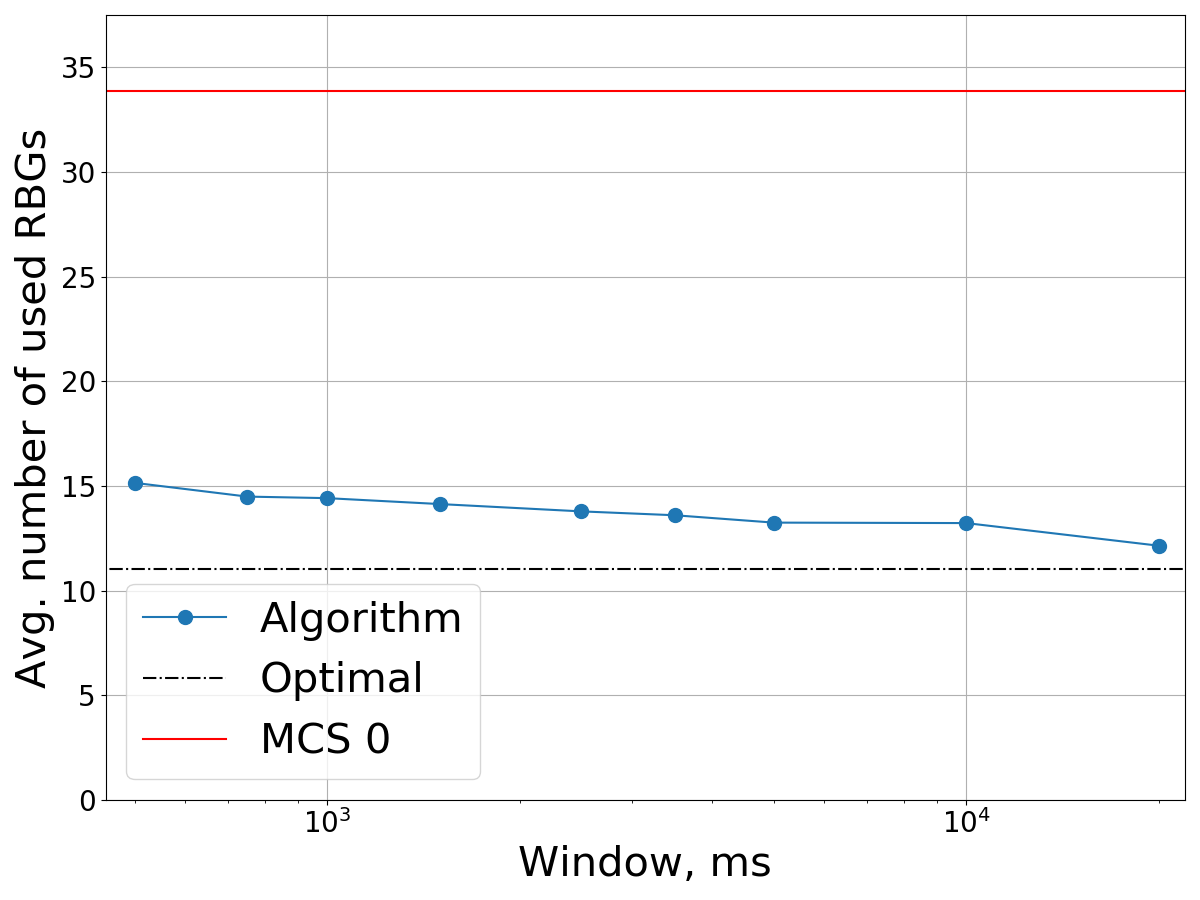}
	\caption{Channel resource consumption of the proposed algorithm.}
	\label{fig:rbu-w}
\end{figure}

Now let us study how the parameters $W$, PLR$_{\text{low}}$ and PLR$_{\text{high}}$ influence the efficiency of the proposed  algorithm. For that, we vary the window size $W$ from 500 ms to 20000 ms and the thresholds PLR$_{\text{low}}$ and PLR$_{\text{high}}$ from $10^{-5}$ to $10^{-15}$, respectively. For each window size value, we find combination of PLR$_\text{low}$ and PLR$_\text{high}$ that provides PLR less than $10^{-5}$ for all SNR$_\text{wb}$ values and the minimum RBG usage averaged over SNR$_\text{wb}$ distribution. Fig.~\ref{fig:rbu-w} shows how the average number of used RBGs depends on the window size for the proposed algorithm. To obtain the optimal and MCS 0 curves, we average the number of used RBGs, presented in Fig.~\ref{fig:opt-msc0}, over the SNR$_\text{wb}$ distribution. According to the results, the number of used RBGs for the proposed algorithm converges to optimal when the window size increases. Moreover, the proposed algorithm allows reducing the resource consumption more than twice in comparison with the fixed selection of the MCS 0.

\begin{figure}[t!]
	\includegraphics[width=0.5\textwidth]{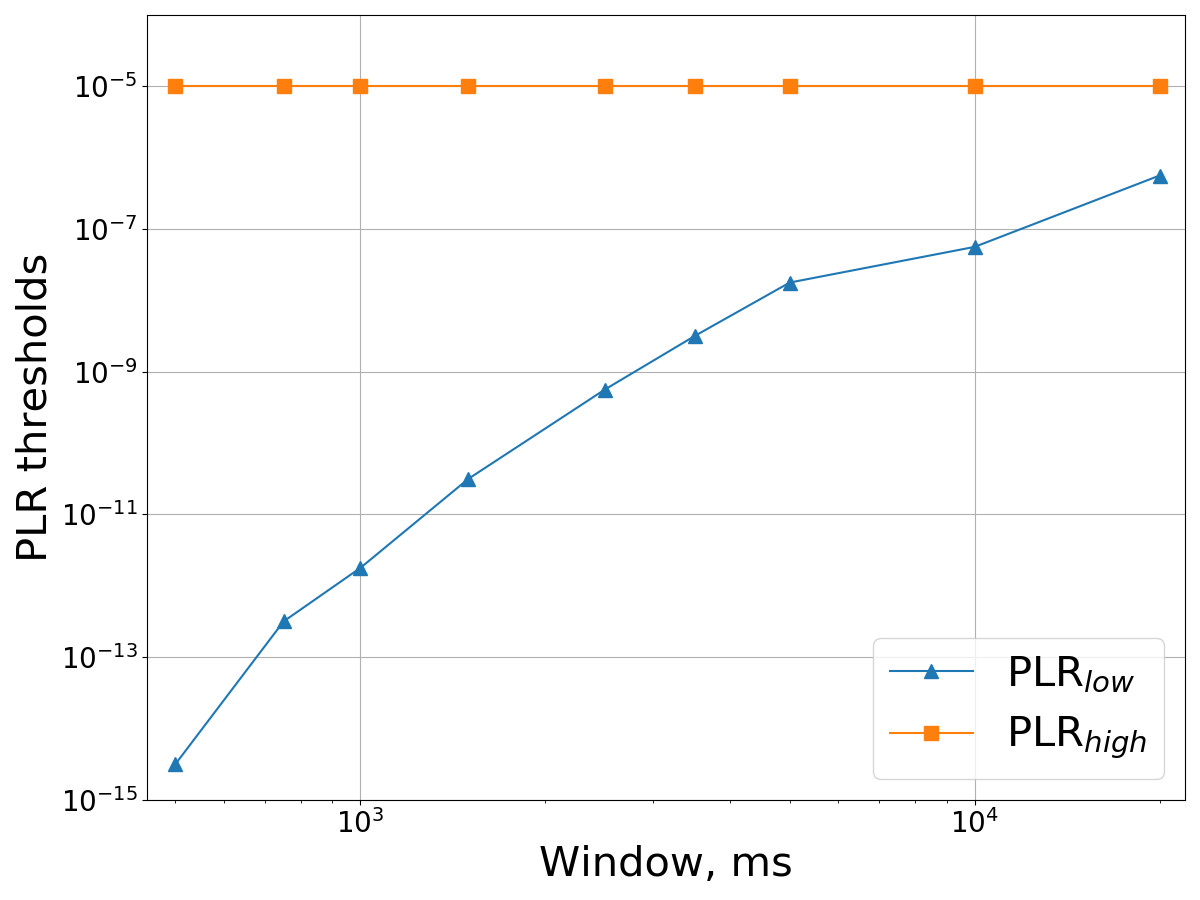}
	\caption{Optimal PLR$_{\text{low}}$ and PLR$_{\text{high}}$ thresholds.}
	\label{fig:plr-w}
	\vspace{-0.3cm}
\end{figure}

Fig.~\ref{fig:plr-w} shows PLR$_{\text{low}}$ and PLR$_{\text{high}}$ thresholds selected by the proposed algorithm. We can see that PLR$_{\text{high}}$ does not depend on the window size and equals $10^{-5}$, which corresponds to $99.999\%$ URLLC reliability requirement. The threshold PLR$_{\text{low}}$ increases with the window size and tends to PLR$_{\text{high}}$ for large window size because the large window allows more conservatively estimating the packet loss ratio provided by different configurations, and thus it does not require to select the lower values for PLR$_{\text{low}}$.

\begin{figure}[t!]
	\includegraphics[width=0.5\textwidth]{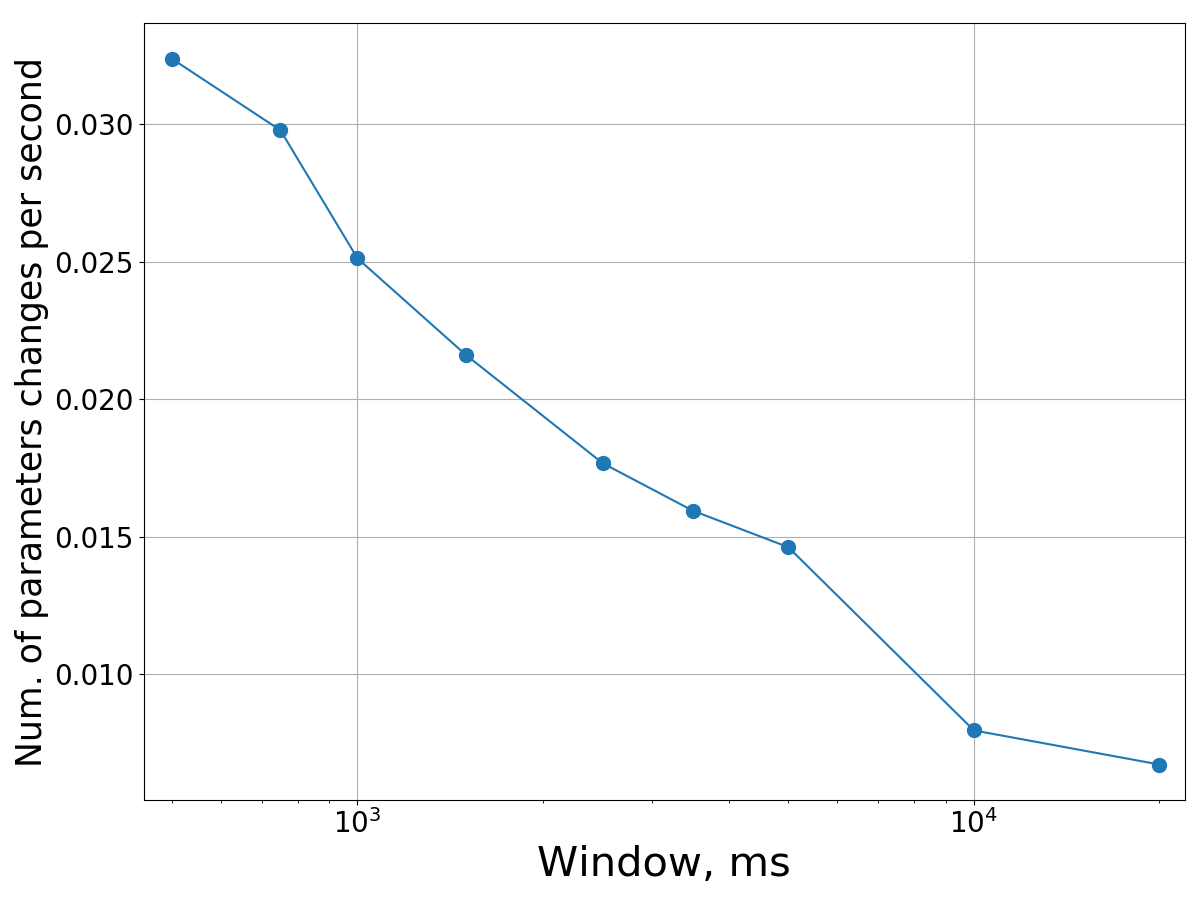}
	\caption{Stability of the selected transmission parameters.}
	\label{fig:switch-w}
	\vspace{-0.4cm}
\end{figure}

According to Fig.~\ref{fig:switch-w}, the usage of two thresholds allows the gNB to rarely reconfigure the transmission parameters. In particular, for $W = 1$s the parameters should be reconfigured once per $40$ seconds on average. The reconfiguration frequency further decreases when the window size increases.

Based on the results presented in Fig.~\ref{fig:rbu-w} and Fig.~\ref{fig:plr-w}, we propose the algorithm to select the parameters of the algorithm. Specifically, we propose to select the window size according to the environmental conditions, e.g., the UE mobility parameters, and then select the PLR$_{\text{low}}$ according to Fig.~\ref{fig:plr-w}. The threshold PLR$_{\text{high}}$ should be set in accordance with the URLLC reliability requirement.

\section{Conclusion}
\label{sec:conclusion}

In this work, we study the problem of uplink transmission parameter selection with the grant-free channel access. We propose the adaptive algorithm for the transmission parameters selection, i.e., MCS and the number of transmission attempts. This algorithm allows satisfying the URLLC requirements while reducing the channel resource consumption. The algorithm uses the signal-to-noise ratio statistics to estimate the packet loss ratio for all possible transmission parameter values. Then the algorithm selects the parameter values that requires the minimum amount of channel resource while satisfying the URLLC requirements. Numerical results obtained with the NS-3 simulator show that the algorithm reduces the channel resource consumption more than twice in comparison with the fixed robust MCS and optimal K parameters selection. Moreover, with this algorithm, the gNB can select the transmission parameters in a long-term perspective, e.g., it can change the parameters for the time intervals much longer than the packet inter-arrival time and, thus, allows reducing the amount of control traffic needed to configure the uplink grant-free transmissions. 

In this paper, we assume that the gNB allocates dedicated channel resources to each UE, and transmission errors are only caused by fading. In our future works, we are going to consider a scenario in which several UEs can use shared grant-free resources to transmit their packets. We will adapt the proposed algorithm to take into account possible interference caused by the transmission of other UEs in shared resources.

\bibliographystyle{IEEEtran}
\bibliography{bibliography}

\end{document}